\documentclass[prd,nofootinbib,showpacs,eqsecnum,amsmath,twocolumn,
titlepage,tightenlines,floatfix]{revtex4}
\usepackage{graphicx}
\usepackage{bm}
\usepackage{dcolumn}
\global\arraycolsep=2pt
\begin{document}

\title{Limits on Production of Magnetic Monopoles Utilizing  Samples from 
the DO and CDF Detectors at the Tevatron}

\date{\today}
\author{G.R. Kalbfleisch}
\email{grk@nhn.ou.edu}
\author{W. Luo}
\altaffiliation{Medical College of Georgia, Augusta, Georgia 30912}
\author{K.A. Milton}
\author{E.H. Smith}
\altaffiliation{Lockheed Martin Advanced Technology Center,
Palo Alto, California 94304}
\author{M.G. Strauss}
\affiliation{Department of Physics and Astronomy, University of Oklahoma, 
Norman, OK 73019}

\begin{abstract}

We present 90\% confidence level
limits on magnetic monopole production at the Fermilab Tevatron from
three sets of samples obtained from the D0 and CDF detectors each exposed to a
proton-antiproton luminosity of $\sim175 \mbox{ pb}^{-1}$ 
(experiment E-882).  Limits are 
obtained for the production cross-sections and masses  for low-mass 
accelerator-produced pointlike Dirac monopoles trapped and bound in material
surrounding the D0 and CDF collision regions.  In the absence of a complete
quantum field theory of magnetic charge, we estimate these limits 
on the basis of a Drell-Yan model.  These results (for magnetic 
charge values of 1, 2, 3, and 6 times the minimum Dirac charge) extend and 
improve previously published bounds.
\end{abstract}
\preprint{OKHEP-03-01}

\maketitle

\section{Introduction}

The existence of even one magnetic monopole with magnetic charge 
$g$ explains  the quantization of electric charge $e$ in 
terms of the Dirac quantization condition \cite{Dirac:1931kp} 
$eg = n \hbar c /2$, $n= \pm1, \pm2, \dots$.  (Throughout this paper
we use Gaussian units, but numerical results are expressed in SI.) 
Besides explaining the quantization 
of electric charge, the existence of magnetic charge
 results in the dual symmetrization of Maxwell's 
equations \cite{Schwinger:1975km}, and is not forbidden by any known 
principles of physics.  
The minimum magnitude of the quantization number is $n=1$ according to Dirac 
or $n=2$
according to Schwinger \cite{Schwinger:1975ww}.  If $e$ is the charge of the 
electron, these magnitudes
 become $n=3, 6$, respectively, if quantization via quark electric charges 
is possible. It should be emphasized that magnetic charge, like electric 
charge, is absolutely conserved, so the lightest magnetically
charged particle is stable, unless annihilated by its antiparticle.

Throughout this paper we refer to magnetically charged particles as magnetic
monopoles, or simply monopoles.  However, as Schwinger emphasized 
\cite{Schwinger:1975km}, magnetically charged particles could also carry
electric charge; such particles he christened dyons.  The quantization
condition for a pair of dyons labeled 1 and 2 is
\begin{equation}
e_1g_2-e_2g_1=n\frac{\hbar c}2.
\end{equation}
We will not explicitly mention dyons further.  We merely note that the
considerations here should  supply similar cross section and mass
limits on dyons as for monopoles.  There would be modifications for
dyons---for example, there would be a binding contribution due to electric
Coulombic attraction or repulsion, but because electric charges are so
much smaller that magnetic ones, the magnetic contributions are overwhelming.
More significant probably are acceptance modifications, due to changes in
energy loss, but we expect the quantitative impact of these changes to be
small (less than 10\%.)

If they exist, monopoles will presumably be abundant, or can be pair-produced
by some appropriate mechanism, and be trapped in matter.
Previous (direct) searches for trapped and bound magnetic monopoles in various
accelerator samples \cite{Eberhard:1975en,Burke:1975px,Giacomelli:1975xy,%
Carrigan:1978ku,Aubert:1983zi,Fryberger:1984fa,Kinoshita:1988cn,%
Kinoshita:1989cb,%
Pinfold:1993mq,Bertani:1990tq,Price:1987py,Price:1990in}, in meteorites 
\cite{Jeon:1995rf}, and lunar soil \cite{Ross:1973it},
 as well as
an earlier result \cite{Kalbfleisch:2000iz,Luo:2002tm} 
from this experiment have been made.
Other (indirect)
searches by other methods are not covered in this paper, but are reviewed
elsewhere \cite{Luo:2002tm,gp}.  Here we report 1) a reanalysis of the data of 
Ref.~\cite{Kalbfleisch:2000iz},
2) the data of Luo \cite{Luo:2002tm} as well as 
3) that of a third set of samples from the
CDF detector recently measured and analyzed.  This extension of limits is
experimentally driven.  Theoretical motivations derive from the expectation that
monopoles from spontaneous electroweak scale symmetry breaking might give rise
to monopoles of mass $\sim2.5$ to $\sim15$ TeV 
\cite{Preskill:1984gd,Kirkman:1981ck}, 
although we here can only search
out to a mass, in our Drell-Yan modeling, $\sim0.4$ TeV.  The LHC using
this method would allow one to approach
2 TeV. Unfortunately no accelerator is currently envisioned that will reach 
the theoretically interesting region of 10 to 15 TeV.  In view of our nearly
complete absence of knowledge of the origin of particle masses, we should 
not exclude any mass region from an experimental search.

This paper presents the experiment, analyses, and results of our search for
monopoles.  Section \ref{secI} 
outlines the basics of the detector apparatus; details 
are provided elsewhere \cite{Luo:2002tm}.  
Section \ref{secII} covers the calibration and linearity 
of the detector.   Section \ref{secIII} 
discusses the samples and monopole energy loss
leading to the stopping and capture of monopoles in the sample material.  Section 
\ref{secIV}
describes the analysis of the measurements.  Section \ref{secV} discusses the
transformation of the data to monopole cross-sections and mass limits.
Finally we summarize in Section \ref{secVI}.

\section{Magnetic Monopole Detector}
\label{secI}

\begin{figure}
\centering
\includegraphics[height=12cm]{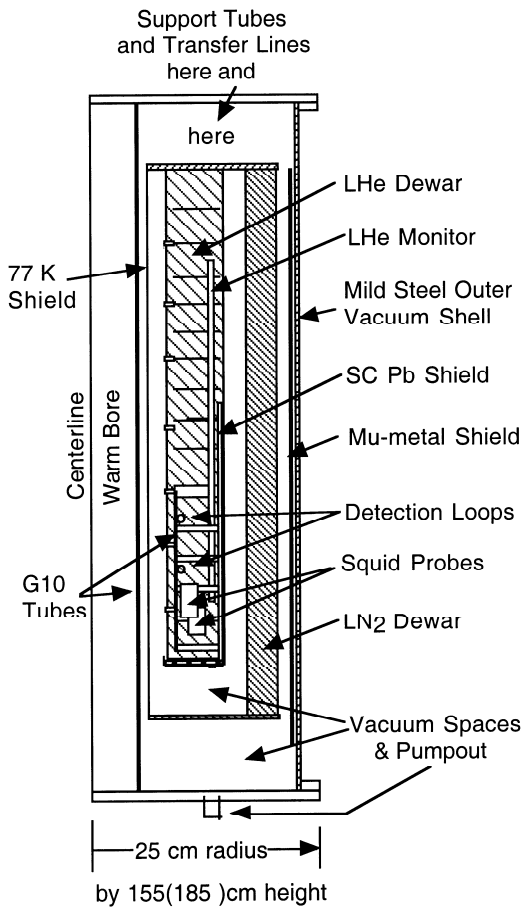}
\caption{Schematic radial cross-section of the monopole detector at a 2:1 
width to height ratio.  The elements generally are rings, tubes, or cylinders
concentric with the indicated centerline.  The height, which  was
155 cm during the set 1 measurements of the D0 aluminum samples, was
increased to 185 cm for the subsequent CDF lead and aluminum samples.}
\label{fig1}
\end{figure}

We use the induction method of Alvarez, et 
al.~\cite{Eberhard:1971re,Alvarez:1963zp} to detect monopoles.  
A large warm bore 
cryogenic detector, similar to that of Jeon and Longo \cite{Jeon:1995rf}, 
was constructed
at the University of Oklahoma.  The active elements of the detector, shown
schematically in Fig.~\ref{fig1}, 
are two 19 cm diameter superconducting loops 
each connected to a dc SQUID (superconducting quantum interference device).
The magnetic flux from a magnetic multipole passing through the loop induces a 
change in the loop's supercurrent because the Meissner effect prevents 
a change in the magnetic flux through such a loop.  If
the multipole is dipole or higher, then the total net change when the 
sample traverses the loop starting and ending at large distances goes 
to zero, although locally supercurrents are present when the sample is in the 
neighborhood of the loop.  But if a monopole passes through the loop, then
a nonzero net change in the supercurrent occurs, giving a current ``step'' 
characteristic of a magnetic charge.  This effect can be observed from even one
monopole in a macroscopic sample due to the long range ($\sim1/r^2$) nature of
the associated magnetic field.   The change in supercurrent is detected
by the SQUID connected to the loop and converted to
a voltage by the SQUID's preamplifier and controller (see below).

Measurement of samples of a size less than 7.5 cm in diameter by 8.5 cm 
in length is made by repeatedly passing them through the 10 cm 
diameter warm bore 
centered on and perpendicular to the loops.  A vertical excursion of 1.1 m
around the position of the loops is typically made.  In a central 65 cm 
region this allows for the magnetic effects of induced and permanent dipole
moments of the sample to start and return to zero on each up and each down
traversal taking some 25 s each.  A net data rate of 10 Hz is recorded for 
each of the SQUIDs.  Also recorded are the readings of an accelerometer, the
vertical position of the sample recorded by an optical encoder, the number 
of increments taken by the stepper motor moving the sample, and the time.
The data acquisition (DAQ) was performed with Apple Macintosh computers running
under National Instruments LABVIEW programs \cite{labview}.

Other magnetic and electronic signals affect the SQUIDs also, causing 
systematic errors.  One has to deal with:
\begin{enumerate}
\item Permanent dipoles in the samples (presumably microscopic
		particles of magnetite or other ferrites)
\item  Induced dipoles (because the samples are conducting metals)
\item  Contamination of the transporting nylon string and copper wire
\item  Ground loops
\item  External electronic device interference (certain clocks,
		welding operations, etc.)
\item  Thunderstorms (which forced suspension of operations)
\item  Mechanical vibrations (external limited by
dampers and isolation, internal due to cryogen boiling)
\item  Small variations in the warm bore magnetic field gradient
\item  Unidentified sources (some days operations had to be suspended).
\end{enumerate}

In addition to data running, background running was required, which was 
subtracted from the data runs.
Even during good running conditions, source (3) was always present to some 
degree.  Source (1) was present for most samples;  large dipole moments off-%
center in the volume of the sample give dipole tails that are unbalanced and
can mimic monopole steps.  Samples with large dipole signals need to be 
vetoed, because the SQUID loses count of the number of flux quanta.
Induced signals, source (2), which are oppositely directed on up/down
traversals, are pairwise cancelled.  The temporal dependence of these
signals [a complete sample measurement takes some twenty (20) minutes]
is minimized by
using only time-coincident traversals from the two squids taking data.
It is to be noted that source (1) is variable.  On repeat traversals, the 
magnitude of the dipole signal may change, most likely because  the 
magnetite grains may be 
re-magnetized.  The induced dipole signal, source (2), is proportional to the
conductivity of the sample, the magnetic field gradient in the warm bore, and
the velocity of the sample.  The velocity used (one meter per 25 seconds) kept 
the magnitude smaller than the permanent dipole value in general.

Sporadic electrical and electronic interferences listed above occasionally 
caused automatic resets of the SQUID controllers.  The operator had to 
restore the output levels of the SQUIDs manually and continue the measurement.
Later, in the analysis stage, these reset (bad) data traces are 
vetoed pairwise.

\section{Calibration and Linearity}
\label{secII}

\begin{figure*}
\centering
\includegraphics[height=15cm,angle=90]{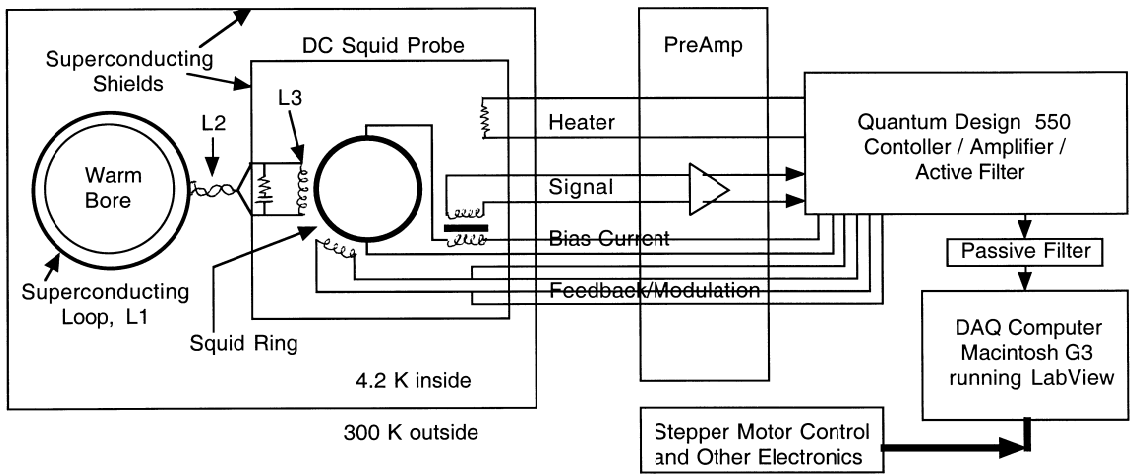}
\caption{Schematic of the monopole detector's signal measuring electronics.
The flux passed through the superconducting loop, L1, is converted into a
current change that induces a flux change in the DC SQUID ring.  An induced
feedback current is amplified and converted into the output signal voltage
in the preamplifier/controller and processed by the LABVIEW data acquisition
(DAQ) program running in the Macintosh computer.}
\label{fig2}
\end{figure*}

The schematic circuitry of the SQUIDs and the associated detection loops
are shown in Fig.~\ref{fig2}.  
During operation of the experiment, the SQUIDs are tuned 
and their transfer functions measured periodically according to the 
manufacturer's specifications to keep them operating with constant sensitivity.
The sensitivity can be approximately predicted from the SQUID sensor's 
parameters and accurately calibrated by measurement. The absolute 
calibration of an expected signal from a Dirac
monopole is made using a ``pseudopole.''  A long thin magnetic solenoid 
carrying a small known current gives a calculable pseudopole (ps) at either
end.  The pseudopole can either be passed through the warm bore of the 
detector in a way similar to the samples (discussed later), 
or it can be placed in a given 
position with one end fully extended through the SQUID loops and the solenoid 
current repeatedly switched on and off.  Both methods were used.  We now
discuss the parameters and the calibration.

The expected response function ($\mbox{RF} = \frac{dV}{d\Phi_{\rm loop}}$)
of the SQUID system 
depends on the following parameters (where the current to voltage conversion
is done by the SQUID controller's amplifiers):
\begin{equation}
\mbox{RF} = \frac{dV}{d\Phi_{\rm squid}}\frac{d\Phi_{\rm squid}}{dI}
\frac{dI}{d\Phi_{\rm loop}}\quad \mbox{in}\quad \mbox{V}/\phi_0,
\label{e1}
\end{equation}
where $dI$ is the change in current induced in the SQUID sensor,
$d\Phi_{\rm squid}$ is the flux change in the SQUID,
$d\Phi_{\rm loop}$ is the flux change in the detection loop,
$\phi_0$ is $hc/2e=2.07\times10^{-15}$ Wb, the superconducting fluxoid unit,
and $dV$ is the change in output voltage of the SQUID controller.
Our SQUIDs were two junction DC types manufactured by Quantum Design 
\cite{C1}.

$dV/d\Phi_{\rm squid}$ is the so-called transfer function, ``XF,''
 measured by a test pulse 
injection procedure given by the manufacturer, for our SQUIDs approximately 
0.5--0.8 $V/\phi_0$. $d\Phi_{\rm squid}/dI$ 
is a parameter measured and furnished
by the manufacturer.
$dI/d\Phi_{\rm loop}$ is calculated as follows. Since the magnetic 
flux linking a superconducting 
loop is conserved as a result of the Meissner effect,
we have by integrating the generalized Maxwell equation
\begin{equation}
- \bm{\nabla}\times\mathbf{ E} = \frac1c\frac{d\mathbf{B}}{dt}  + 
\frac{4\pi}c\mathbf{J}_m,
\label{e2}
\end{equation}
$\mathbf{J}_m$ being the magnetic current density,
the following relation for the current induced in the detection loop
by one flux quantum passing through it (see Appendix):
\begin{equation}
I =\frac{\phi_0}c\frac{\left(1-\left(\frac{r}{a}
\right)^2\right)}{L_1 + L_2 + L_3},
\label{e4}
\end{equation}
where (see Fig.~\ref{fig2}) $L_1$ is the inductance of the detection loop,
$L_2$ is the inductance of connecting twisted pair,
$L_3$ is the input inductance of the squid sensor,
$r$ is the radius of detection loop, and
$a$ is the radius of the superconducting lead (Pb) shield.
The factor $( 1 - (r/a)^2 )$ inserted in Eq.~(\ref{e4})
corrects for the trapping of flux coupled into the 
detection loop due to the cylindrical superconducting shield. (See Appendix.)

The sensor's 
input coil inductance $L_3$ is measured to be of order 1.8 $\mu$H (see Table
\ref{tabI}).
The inductance $L_2$ is zero, and $L_1$ (of order 1 $\mu$H) 
is calculated from the standard formula \cite{CE,stratton}:
\begin{equation}
L_{\rm loop}(C) = \frac{4 \pi}{c^2} r \left( \ln \frac{8r}{\rho} -C\right),
\label{e5}
\end{equation}
where $r$ is the radius of the loop and $\rho$ is the wire radius. For a 
uniform current density in the loop $C=7/4$; for a (superconducting) surface 
current $C=2$ (a reduction in the inductance
of 4 percent).  In addition, the superconducting 
shield reduces the value of $L_{\rm loop}$ by 9.5 percent:
\begin{equation}
L_1= L_{\rm loop}(2)(1 -0.095) = 0.70 \mu\mbox{H}.
\label{e6}
\end{equation}
This comes about as follows: The inductance between the detection loop and
the superconducting shield is given by the following formula, rather
easily derived from the inductance between two loops \cite{stratton}:
\begin{widetext}
\begin{equation}
L_{ls}=\frac{4\pi}{c^2}  \int_0^\pi d\phi\,\sin^2\phi\frac{a^2r^2}
{(a^2+r^2+2ar\cos\phi)(a^2+r^2+l^2+2ar\cos\phi)^{1/2}},
\label{e7}
\end{equation}
\end{widetext}
where $a$ is the radius of the shield and $2l$ its length.  
Inverting the inductance matrix changes the effective
inductance of the detection loop by a significant amount
\begin{equation}
L_1 = L_{\rm loop} \left( 1 - \frac{L_{ls}^2}{L_{\rm loop}L_{\rm shield}} 
\right),
\label{e8}
\end{equation} and putting in the values 
$a = 14.73$ cm, $r = 9.855$ cm, $l = 30$ cm, and $\rho = 0.02$ cm 
gives $L_1$ as in Eq.~(\ref{e6}), 
where $L_{\rm shield} = 4\pi a^2/(2lc^2)$ is the self-inductance of 
the superconducting cylindrical shield.

The experiment had four (4) SQUID sensors referred to as
 DC1, DC2, DC3, and DC4.
DC1 failed at some early time, was repaired and subsequently called DC1R.
Most of the sample measurements for the final analysis were
performed with the pair DC1R and DC2.  However,  some measurements were 
made with other pairings
of the four SQUIDs, and in a number of configurations, which
are catalogued in the tables which follow.  In initial tests, we had
an emulation setup with a small bore and small detection loops immersed in 
liquid helium contained in a ``Research Dewar.''  We also had two current 
sources for the pseudopole, one having multiples of a current unit 
[$\sim1.55$ V battery/500 M$\Omega$, equal to about 0.7 Dirac poles---see
Eq.~(\ref{e10}) below] equal 
to 1, 2, 5, 25, 100 current 
units and a later one with multiples of 1.7, 2.5, 5, 50 
current units.  We also had an early pseudopole 
of 0.5 meter length and a later one of 1.016 meter length.  The actual currents
for each number of units was measured with a picoammeter to be correlated
with the voltage response of the calibration measurements.

\begin{figure*}
\centering
\includegraphics[height=10cm]{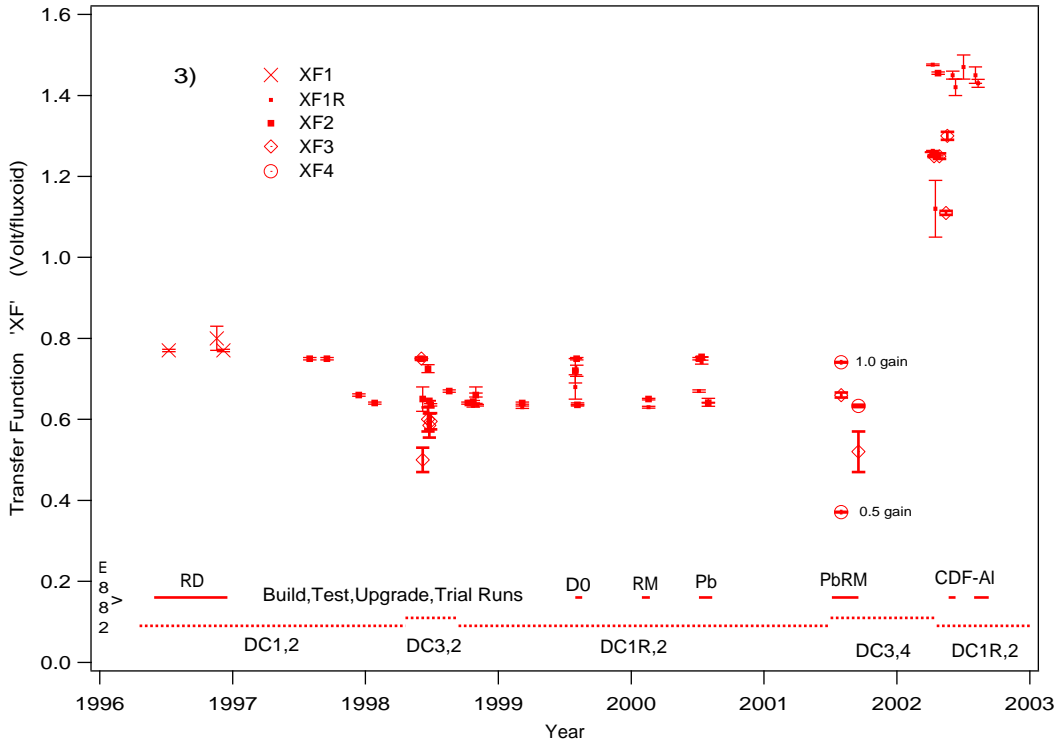}
\caption{The myriad of transfer function (XF $=dV/d\Phi_{\rm squid}$) 
measurements made over the 
course of the monopole experiment.  Time intervals for the SQUID
configurations in use at the various times, as well as the periods of
measurement of the different sample sets are also indicated. RD $=$ research
dewar, D0 and CDF are the collider experiments from which the samples
were taken, RM $ = $ remeasurements and 1, 2, 3, 4 
refer to the various SQUID probes 
(see text).}
\label{fig3}
\end{figure*}

The individual measured transfer functions, XF, in $V/\phi_0$
 are given in Fig.~\ref{fig3}.
We see two lines of values at 0.75 and 0.64 for most of the plot,
with values of 1.5 and 1.3 at year 2002.5, where we doubled the output gain
of the squid controllers.  The double valuedness is due to the presence
or absence of an external passive RC low pass filter, which exhibits
attenuation to the signal before being converted on the ADC board in the
computer.  (The controller has an internal active low pass filter which can
be switched on or off, which in comparison is attenuation-free well below 
the rolloff point.)  Table \ref{tabI} identifies in the name DCi the filter 
configuration as ``NF'' or ``EF'' for ``no filter'' (or non-attenuating active 
internal filter) or ``external filter,'' respectively.  
In the table, *2 or RD indicate the doubled gain or early Research Dewar data. 
A gain 0.5 data point was taken while testing modifications to the controllers,
and is exactly half of the gain 1.0 point above it (see Fig.~\ref{fig3}).  
The transfer functions and their errors shown in Fig.~\ref{fig3} 
have been averaged 
as appropriate and entered into Table \ref{tabI}.  
Table \ref{tabI} also shows the $d\Phi_{\rm squid}/dI$ 
values, the input coil inductances $L_3$, the
$dI/d\Phi_{\rm loop}$ values calculated from Eq.~(\ref{e4}), 
and the theoretical RF's [Eq.~(\ref{e1})] shown
(mV/Dirac pole).  Note one Dirac pole corresponds to two fluxoid units,
$4\pi g=h c/e$.

\begin{table*}
\caption{\label{tabI}Transfer functions and predictions. The subscript
$s$ denotes SQUID.  Here $\{L_1, 1-(r/a)^2\}=\{0.70\, \mu\mbox{H}, 0.552\}$
for all entries except for the first (RD) where they are $\{0.01\,\mu\mbox{H},
0.87\}$.  Remember $\phi_D=2\Phi_0$.
$\Delta (dV/d\Phi_s)$ is the error in $dV/d\Phi_s$, and likewise 
$\Delta({\rm RF}_{\rm th})$ is the error in ${\rm RF}_{\rm th}$.}
\begin{ruledtabular}
\begin{tabular}{rcccccccc}
Point&SQUID/Filter&$dV/d\Phi_s$&$\Delta (dV/d\Phi_s)$&$d\Phi_s/dI$&$L_3$
&$dI/d\Phi_{\rm loop}$&
${\rm RF}_{\rm th}$&$\Delta({\rm RF}_{\rm th})$\\
&&\multicolumn{2}{c}{(V/$\phi_0$)}&($\phi_0$/$\mu$A)&($\mu$H)&(nA/$\phi_0$)
&\multicolumn{2}{c}{(mV/$\phi_D$)}\\
0&DC1NFRD&0.770&0.002&5.24&1.88&0.953&7.69&0.15\\
1&DC1NF&0.770&0.002&5.24&1.88&0.442&3.56&0.08\\
2&DC1RNF&0.739&0.005&5.24&1.85&0.447&3.46&0.08\\
3&DC1REF&0.637&0.004&5.24&1.85&0.447&2.98&0.07\\
4&DC2NF&0.752&0.001&4.17&1.47&0.525&3.29&0.09\\
5&DC2EF&0.643&0.002&4.17&1.47&0.525&2.82&0.07\\
6&DC3NF&0.750&0.003&5.24&1.85&0.447&3.51&0.08\\
7&DC3EF&0.643&0.015&5.24&1.85&0.447&3.01&0.10\\
8&DC4NF&0.741&0.001&3.94&1.43&0.535&3.12&0.08\\
9&DC4EF&0.633&0.003&3.94&1.43&0.535&2.67&0.07\\
10&DC1RNF*2&1.473&0.005&5.24&1.85&0.447&6.90&0.15\\
11&DC1REF*2&1.260&0.002&5.24&1.85&0.447&5.90&0.13\\
12&DC2NF*2&1.455&0.003&4.17&1.47&0.525&6.37&0.17\\
13&DC3NF*2&1.252&0.005&5.24&1.85&0.447&5.86&0.13\\
14&DC3EF*2&1.110&0.005&5.24&1.85&0.447&5.20&0.12\\
\end{tabular}
\end{ruledtabular}
\end{table*}

\begin{figure}
\centering
\includegraphics[height=9cm]{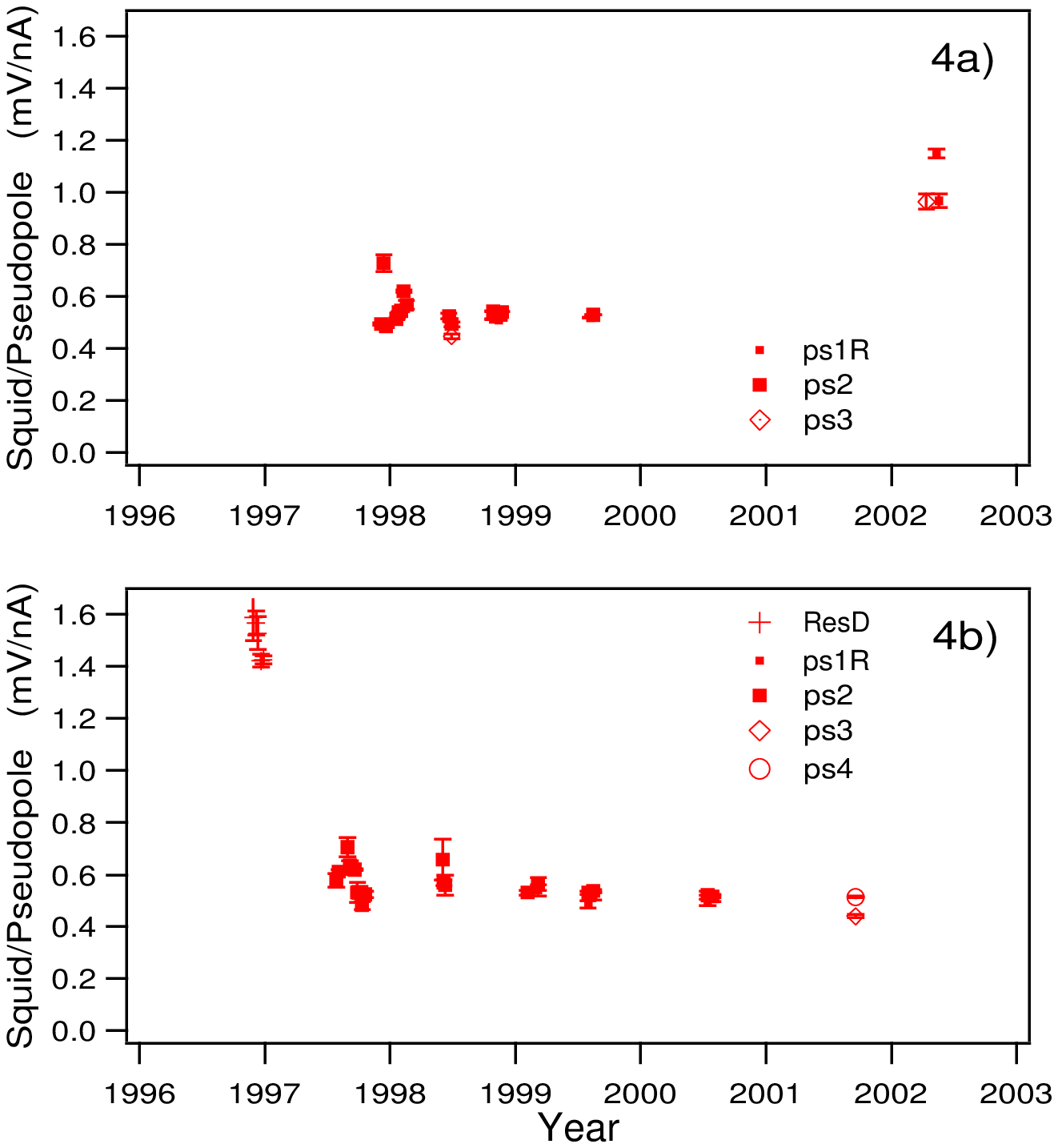}
\caption{Individual measured pseudopole responses (mV/nA) for a) traversing
the loop similarly as the samples and b) turning the current on/off while
the pseudopole is ``parked'' at one position.}
\label{fig4}
\end{figure}

The experimental data (mV/nA) for traversals of the pseudopole through the 
warm bore,
as with samples, and for the pseudopole ``parked'' at one position and its
current turned on and off, are shown in Fig.~\ref{fig4}a and \ref{fig4}b,
 respectively.  The two methods agree within experimental uncertainties.
 One again
sees (at early times) the RD data, then the normal mid-year data, and 
finally the doubled gain data in 2002.  Averaging 
the appropriate groupings of DCi gives the values (mV/nA) in 
Table \ref{tabII}.  
\begin{table*}
\caption{\label{tabII}Measured pseudopole
responses and comparisons to predictions (Table \ref{tabI}).
(Note:  mV/nA $\times$ nA/$\phi_D$ = mV/$\phi_D$.)
Here ``Ratio'' is the ratio of the theoretical response RF$_{\rm th}$
given in Table \ref{tabI} to the measured response/pole.}
\begin{ruledtabular}
\begin{tabular}{rccccccc}
Point&SQUID/Filter&Response&$\Delta$(Response)
&$i$/pole&Response/pole&$\Delta$(Response/pole)&Ratio\\
&&\multicolumn{2}{c}{(mV/nA)}&(nA/$\phi_D$)&
\multicolumn{2}{c}{(mV/$\phi_D$)}&\\
0&DC1NFRD&1.44&0.02&4.40&6.34&0.10&1.21\\
1&DC1NF&&&4.63&&&\\
2&DC1RNF&&&4.63&&&\\
3&DC1REF&0.51&0.01&4.63&2.36&0.05&1.26\\
4&DC2NF&0.62&0.003&4.63&2.87&0.02&1.14\\
5&DC2EF&0.535&0.002&4.63&2.48&0.02&1.14\\
6&DC3NF&&&4.63&&&\\
7&DC3EF&0.444&0.014&4.63&2.06&0.07&1.46\\
8&DC4NF&&&4.63&&&\\
9&DC4EF&0.513&0.003&4.63&2.38&0.02&1.12\\
10&DC1RNF*2&1.15&0.01&4.63&5.32&0.06&1.30\\
11&DC1REF*2&0.97&0.013&4.63&4.49&0.07&1.31\\
12&DC2NF*2&&&4.63&&&\\
13&DC3NF*2&&&4.63&&&\\
14&DC3EF*2&0.97&0.016&4.63&4.50&0.08&1.15\\
\end{tabular}\end{ruledtabular}
\end{table*}

The conversion from the response in mV/nA to that in mV/Dirac pole
is given by
considering the pseudopole solenoid's magnetic moment both as a current
loop and as a pole-antipole dipole, i.e.,
\begin{equation}
	g L  = \frac1c N i A
\label{e9}
\end{equation}
where
the length of solenoid is $L = 1.016 \,(0.50)$ m for the new (old) pseudopole, 
respectively,
$Ni$ is the ampere-turns of current, $N = 4710\, (2440)$ turns,
and the cross-sectional area of 
the solenoid is $A = 1.533 \pm0.011 \times 10^{-4} \,
\mbox{m}^2$,
so that (in SI, $g_D/\mu_0=3.29\times 10^{-9}$ A\,m)
\begin{equation}
\frac{i}{\rm pole} =\frac{ gc} { (N/L) A } = 4.63 (4.40) \pm0.03\,
\mbox{nA}/\phi_D,
\label{e10}
\end{equation}
for the new (old) pseudopole, respectively.

\begin{figure}
\centering
\includegraphics[height=8cm]{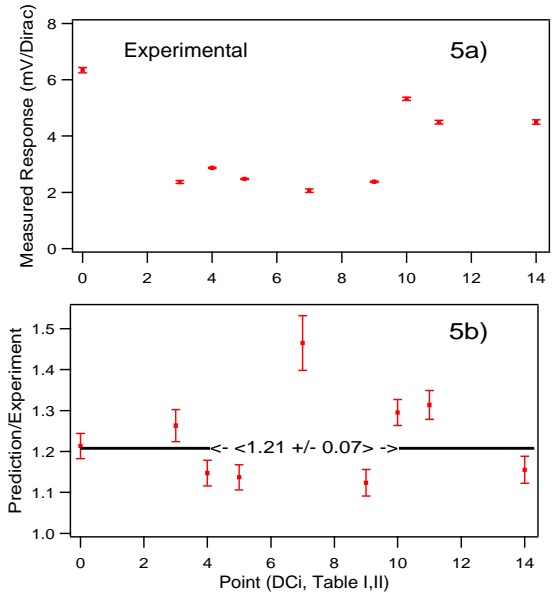}
\caption{Experimental calibration of squid responses (a) as compared to
predictions (b).  The values from (a) are used in the data analysis of the
samples below.  The identity of the respective squid is indicated by ``Point.''
}
\label{fig5}
\end{figure}

\begin{figure}
\centering
\includegraphics[height=9cm]{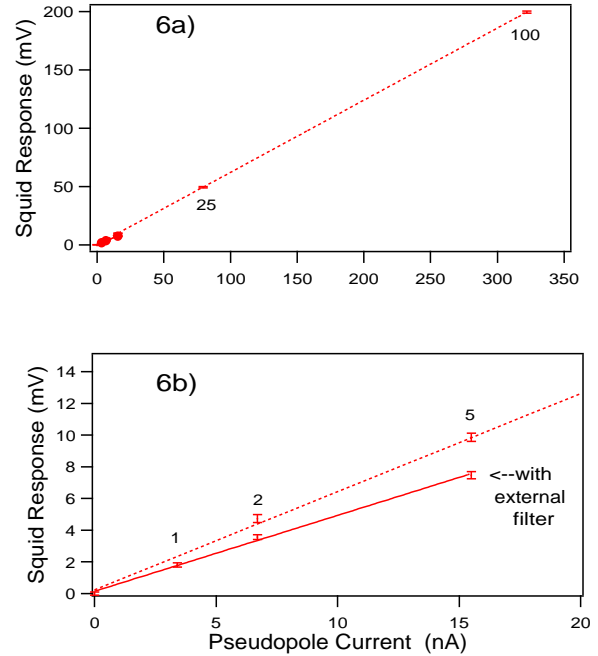}
\caption{Linearity for no filter and for passive filter for a) the  whole range
of pseudopole currents and b) expanded.  The units of current (about
0.7 Dirac pole, see 
text) are indicated.  This is data from DC2 (1997) for the short pseudopole.
}
\label{fig6}
\end{figure}

Using these conversion values we get the measured 
pseudopole response in mV/$\phi_D$
pseudopole values versus DCi configuration shown in Table \ref{tabII} and
Fig.~\ref{fig5}a and the expected/measured ratio in Table \ref{tabII}
 and Fig.~\ref{fig5}b, respectively.
 A weighted average of these ratios for the nine values is 
$1.21 \pm 0.07$.
The discrepancy is presumably due to approximate values of Quantum Design's
measured parameters, and the elementary treatment excluding shielding internal
to the SQUID probes, etc.
The values plotted in Fig.~\ref{fig5}a are used to determine cuts on the 
possible monopole ``steps'' seen in the analysis
of the data given later below.  Data taken at year 1997.7 give the linearity
response shown in Fig.~\ref{fig6}a and \ref{fig6}b.  
One sees that the calibration is very 
linear, and that signals down to one-half a Dirac pole can, in practice,
be measured.

\begin{figure}
\centering
\includegraphics[height=13.5cm]{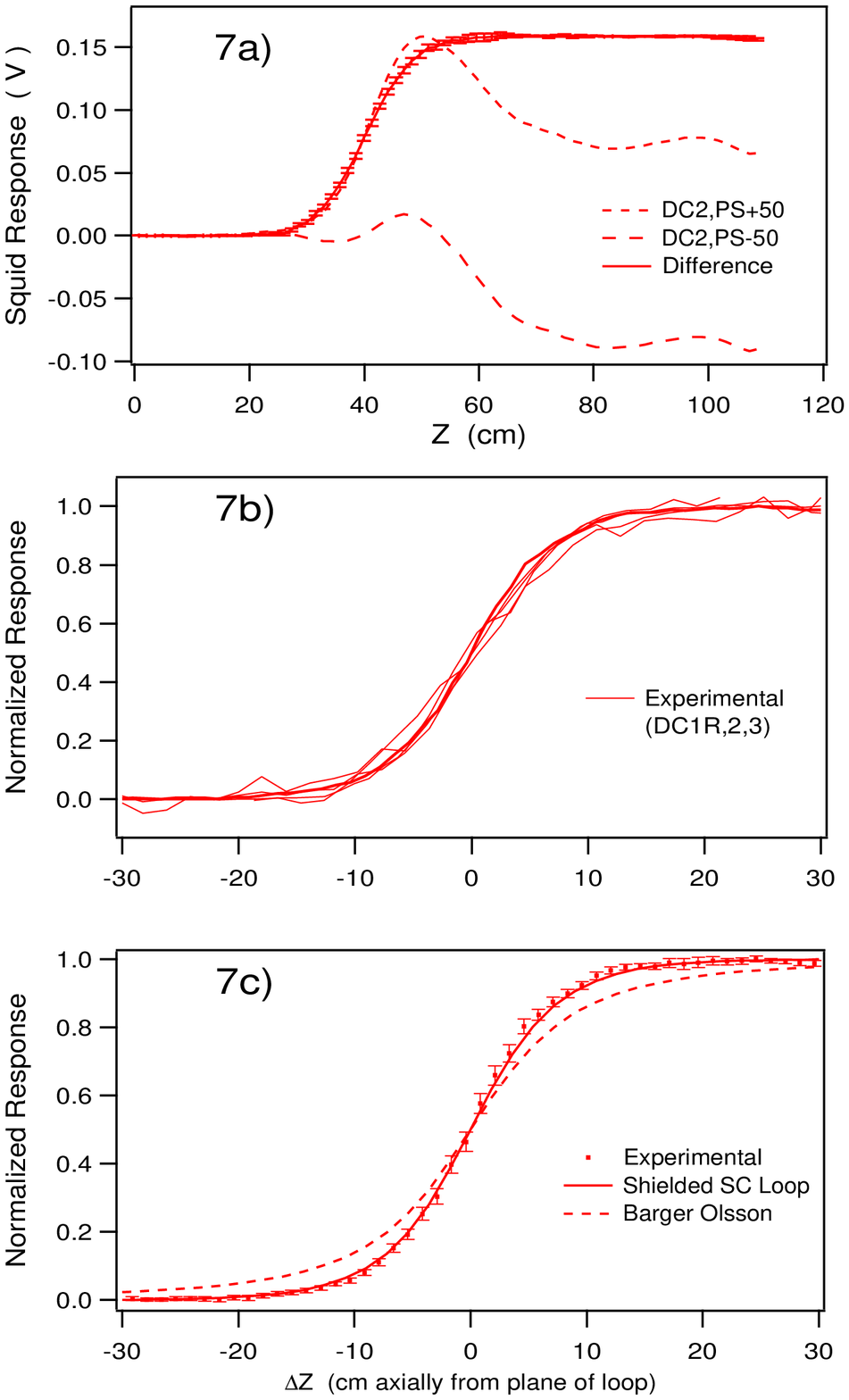}
\caption{ a) Pseudopole response, SQUID DC2, versus position for two currents,
$\pm 50$ current units (about $\pm 35$ Dirac poles) 
and their difference, giving
the experimental monopole step shape, and b) several such shapes [one of DC1R,
three of DC2 including 7a) above, and one of DC3] compared,
and c) Fig.~\ref{fig7}a compared to the theoretically
expected shapes: loop in free space (poor) \cite{bargerandollson}
and in a superconducting
can (SC) (good)--see Appendix.}
\label{fig7}
\end{figure}

The shape of an expected monopole step is obtained by a subtraction of 
two runs with different pseudopole currents, as shown in Fig.~\ref{fig7}a.
Five such shapes are compared in Fig.~\ref{fig7}b.
The theoretical modeling, given in the  Appendix, 
is compared to the experimental 
data in Fig.~\ref{fig7}c.  
The agreement between the theoretical and experimental shapes is good.  
We conclude that the calibration, linearity, and positional response (shape) 
indicate that the experiment is understood, and that operating conditions 
over six years are reasonably consistent and stable.

\section{Samples, Stopping and Trapping}
\label{secIII}

\subsection{Samples}
\label{subsecA}

There were three sets of samples obtained from discarded material from the
upgrading of the  
D0 \cite{d0} and CDF \cite{cdf} detectors: 1) Be beam pipe and Al ``extension''
 cylinders from D0,
2) Pb from the forward/backward ``FEM'' (Forward ElectroMagnetic calorimeters)
of CDF, and 3) half of the Al cylinder (``CTC'' support) from CDF.  Sample set
1 was initially reported \cite{Kalbfleisch:2000iz}, 2 was given in Luo's PhD. 
thesis \cite{Luo:2002tm}, and
set 3 here.  In this paper, all three sets are being reported according to a
final consistent analysis.

Sample set 1 comes from the two Al extension cylinders (extending beyond the
main central detectors of D0 inside the liquid argon calorimetry), each of 150 
cm diameter by 46 cm length by 1.26 cm thickness.  They came as 16 plates.
Four of these plates were further cut by water jet, whereas the balance were
cut on a bandsaw; no magnetic degradation was observed from the bandsawing
in comparison to the water jet samples.
These cut pieces were 7 cm by 7.6 cm or 6 cm by
7.6 cm in size.  Two of each were bundled into a ``cylinder'' constituting
a ``sample.'' There are a total of 222 Al samples.  The Al sawings and other 
small scrap pieces are accounted for later.

In addition, sample set 1 included the D0 5 cm diameter, 0.05 cm thickness
 Be beam pipe.  The
central 46 cm section, centered on the collision region and covering nearly 
the full solid angle, was cut into six 7.6 cm pieces.

\begin{figure}
\centering
\includegraphics[height=7cm,angle=-181]{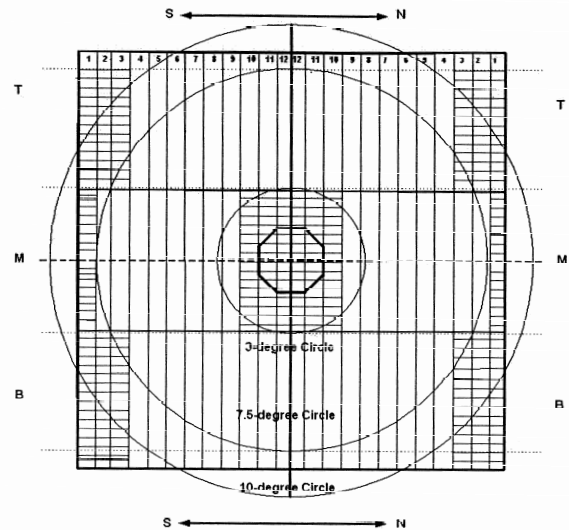}
\caption{A lead sample layer is cut into 68 samples. The Monte Carlo
simulation shows that only the samples (52 in one layer) subtending an
angle $3^\circ < \theta < 7.5^\circ$ have significant acceptance and should be
considered.  The cross-hatched samples and others with cuts on large $dS$
(statistical) and $sS$ (systematic)
errors are excluded from the analysis; the (solid angle) coverage 
falls to 53\% for $n=1$ monopoles (see text).
}
\label{fig8}
\end{figure}

Sample set 2 comes from twelve 2 m by 2 m
by 0.5 cm thick Pb sections cut from the center of the full 3.04 m by
3.04 m Pb layers.  These twelve layers were those closest to the interaction 
region of which six were located on the east side and the other six on 
the west.  Each layer had an octagonal hole at the center (of approximately 
30 cm ``diameter'') for the passage of the $\overline p$-$p$ beams, etc. 
Each 2 m by 2 m 
section came as six pieces of 0.667 m by 1 m, labeled T (top), M (middle) or
B (bottom) and also labeled N (north) or S (south).  The six sections were
also labeled as E (east) or W (west) and by layer number (1,2\dots,6).  These
sections were sheared
into twelve strips of 8.3 cm width across the
1.00 m dimension, and each strip was rolled
into a cylinder of approximately 7.5 cm 
diameter.  These constitute some 816 samples.  A typical layer is shown 
in Fig.~\ref{fig8}.  

Of the 816 samples, only 664 have been successfully measured.  The rest were 
unmeasurable due to huge dipole moments caused by the (ferromagnetic) 
red paint on the first layer samples.  Fortunately, layer 1 samples
turn out not to be needed. The large solenoidal magnetic field of CDF
accelerates the monopoles to a high enough energy (70$n$ GeV additional) 
that they penetrate the early layers, reaching layers 4, 5, 6 for $n=1$ and 
layer 2 for $n= 2,3$, and not at all for $n= 6$.  The scrap pieces and other 
unmeasurable samples of Pb are accounted for later.

Sample set 3 comes from the CDF's inner support cylinder, which has eight
0.63 cm thick cover plates on a cylinder of 2.74 m diameter by 2.64 m length.
Six of these cover plates, covering three-quarters
 of the azimuthal angle about the
beam direction, were made into samples, similar to the D0 ones.
These were sheared on a large shearing machine, first into long strips and 
then each strip into shorter sections.
Eight pieces (four of each size) were bundled into a 7.5 cm diameter by 7.6 cm
long cylinder (the CDF Al being half the thickness of the D0 extension pieces)
yielding 404 samples with little scrap.  Due to limitations on time, funding
and personnel, 132 of the samples did not get measured,
leading to only one-half the azimuthal solid angle being covered.  These 
and the scrap are accounted for later below.

It turns out that all of the CDF Al samples were very magnetic (similarly to
the CDF Pb red painted layer 1 samples).  We had to demagnetize them.  
Degaussing with an AC field coil was ineffective.  But as ferrites have
Curie points below 585 $^\circ$C \cite{kittel} and the melting point of 
Al is 660 $^\circ$C 
(as opposed to 327 $^\circ$C for Pb), we
were able to demagnetize all but a few of the samples by a heat treatment,
``soaking'' the samples at 610 $^\circ$C for one hour.\footnote{The
resulting small increase in thermal energy is completely negligible
compared to the binding energy of monopoles to matter.
Monopoles bound with a keV of
more of energy should be permanently trapped in the material.  See
Ref.~\cite{binding}.}
  The dipole amplitudes
decreased from an unmanageable 3--10 V to the usual 30--100 mV.

\subsection{Stopping of the Monopoles}
\label{subsecB}
\subsubsection{Energy Loss}
\label{subsubsecEL}

\begin{figure}
\centering
\includegraphics[height=10cm]{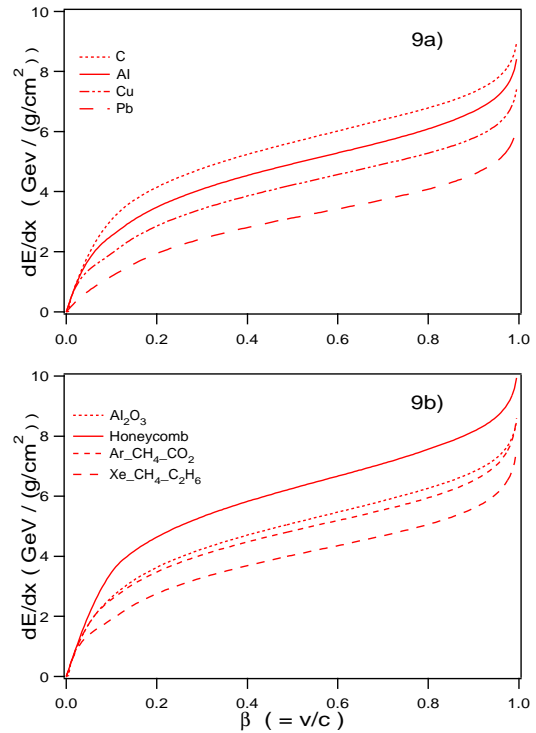}
\caption{$dE/dx$ vs $\beta$ ($=v/c$) for a) elements and b) compounds and 
mixtures
involved in the energy losses of monopoles passing through the various
detector materials. To increase legibility of the elemental figure, we have not
shown the energy loss for Si and Be, which are similar to Al,  nor
for N$_2$, which is similar to C.   For the compounds and mixtures, we
note that C$_3$H$_6$, CO$_2$--C$_2$H$_6$--Al,
 and Rohacell are similar to Honeycomb, and SiO$_2$
is similar to Al$_2$O$_3$. These $dE/dx$ curves are for a 
magnetic charge value of $n=1$; apart from the correction terms $K(|n|)$
and $B(|n|)$, we multiply by $n^2$ for larger magnetic charge 
values.
}
\label{fig9}
\end{figure}

The energy losses ($dE/dx$) of magnetic monopoles traversing material absorbers
are caused by the interaction of the moving monopole charge 
($g=ng_{D}$) 
with the electric field of the atomic electrons,
i.e., the $g {\bf v \times E}$ Lorentz force.  
The velocity dependence of this force cancels
the 1/(velocity)$^2$ dependence of the usual charged particle $dE/dx$. Either 
classically \cite{CE}, 
quantum mechanically \cite{Schwinger:1976fr}, 
or field theoretically \cite{Gamberg:1999hq}, approximately 
one simply substitutes $(g\beta)^2$ for $(ze)^2$ in the usual charged particle
$dE/dx$ formula.  Kazama, Yang, and Goldhaber \cite{Kazama:1977fm} have obtained the
differential scattering cross section for an electron moving in the magnetic 
field of a fixed magnetic pole. Ahlen 
\cite{Ahlen:1982rw,Ahlen:1978jy} 
has used this cross-section to 
obtain the following expression for monopole stopping power:
\begin{eqnarray}
-\frac{dE}{dx}&=&\frac{4\pi}{c^2}
\frac{g^2e^2}{ m_e}N_e\bigg(\ln\frac{2m_ec^2\beta^2\gamma^2}
{I}+\frac12K(|n|)\nonumber\\
&&\quad\mbox{}-\frac12\delta-\frac12-B(|n|)\bigg),
\label{e11}
\end{eqnarray}
where $N_e$ is the number density of electrons, $I$ is the 
mean ionization energy,
 $K(|n|)= 0.406$ (0.346) is the Kazama, Yang and Goldhaber correction 
for magnetic charge  $n=1$ ($n\ge2$) respectively, $\delta$ is the usual density
correction and  $B(|n|)= 0.248$ (0.672, 1.022, 1.685) is the Bloch
correction for $n=1$ ($n= 2, 3, 6$), respectively \cite{Hagiwara:2002fs}.  
(Of course, one must divide by the density to get $dx$ in g/cm$^2$.)
 This formula is good only for velocities $\beta = v/c \agt 0.1$.
For velocities $\beta \alt 0.01$, we use Eq.~(60) of Ref.~\cite{Ahlen:1982mx}
as an approximation for all materials:
\begin{equation}
-\frac{dE}{dx} = (45\mbox{ Gev/cm})  n^2 \beta, 
\label{e13}
\end{equation}
which is linear in $\beta$ in this region.  
The two $dE/dx$ velocity
regions are
joined by an empirically fitted polynomial in the region of $\beta = 0.01$--0.1
in order
to have a smooth function of $\beta$. For the elemental and composite materials
found in the D0 and CDF detectors, we show the resulting $dE/dx$ curves 
we used in Fig.~\ref{fig9}.  (See Ref.~\cite{Luo:2002tm}.)
\subsubsection{Stopping of Possible Monopoles}
\label{subsubsecStop}

The trajectories of possible monopoles are generated in a Monte Carlo 
program from the collision point through the appropriate detector elements 
slowing to a stop in the sample layers, including the
acceleration (or deceleration) of the monopole along the external 
magnetic field lines
of the CDF detector. The average polar angle $\theta$ relative to the beam 
direction is different for the three
sample sets: $\theta= 35^\circ$, $6.5^\circ$, and $90^\circ$ for sets 1, 2,
and 3, respectively.
The amount of material along a ``typical'' trajectory is 
shown in Table \ref{tabIII} for D0 and Table \ref{tabIV} for CDF.
\begin{table*}
\caption{\label{tabIII} The materials and path lengths at an angle of 
$35^\circ$ relative to the beam direction
 for the D0 aluminum sample measurements (set 1).}
\begin{ruledtabular}
\begin{tabular}{ccc}
Detector&Material&Path lengths (g/cm$^2$)\\
&Be pipe&0.16\\
VTX&C$_2$H$_6$ and CO$_2$ gas, Al wire mixture&0.12\\
TRD&Al window&2.18\\
&N$_2$ gas&0.01\\
&Polypropylene&1.12\\
&Xe, CH$_4$, C$_2$H$_6$ gases&0.02\\
&Honeycomb&1.99\\
CDC&Ar, CO$_2$, CH$_4$ gases&0.07\\
&Kapton&0.1\\
&Rohacell&0.49\\
&Endplate (Al)&4.18\\
&G10&1.11\\
&&$\hrulefill$\\
Total including&Sample layer (Al)&11.6--17.6
\end{tabular}
\end{ruledtabular}
\end{table*}

\begin{table*}
\caption{\label{tabIV}
 The materials and path lengths in the CDF detector, at an angle
of $6.5^\circ$ from the beam for CDF lead samples (set 2), and at an angle of 
$90^\circ$ from the beam for the CDF aluminum samples (set 3).  The Pb samples
are spaced by 0.6 g/cm$^2$ of drift chambers.}
\begin{ruledtabular}
\begin{tabular}{cccc}
&&Set 2&Set 3\\
Detector&Material&Path lengths (g/cm$^2$) at $6.5^\circ$&
Path lengths (g/cm$^2$) at $90^\circ$\\ 
&Be pipe&0.82&0.092\\
SVX&C&---&0.209\\
&Al&---&0.135\\
&Si&---&0.210\\
&Al$_2$O$_3$&0.29&---\\
VTPC&C&3.57&0.312\\
FTC/CTC&Al&FTC:\; 0.25&CTC:\; 0.293\\
&Pb layer and drift chamber&6.5\\
&&$\hrulefill$&\hrulefill\\
Total including&Sample layers&Pb layer 2: 11.4--17.2&Al: 1.25--2.97\\
&&Pb layer 4: 24.5--30.2&\\
&&Pb layer 5: 31.1--36.8&\\
&&Pb layer 6: 37.6--43.3&\\
\end{tabular}
\end{ruledtabular}
\end{table*}

\subsubsection{Trapping of Monopoles}
\label{subsubsecTrap}
The monopoles having been stopped can then bind to the magnetic moments
of the nuclei of the material present, and are therefore trapped. 
 The interaction of the monopoles with
the magnetic moments of nuclei and electrons can be strong enough to produce
bound states under certain conditions and be trapped, having a very long 
lifetime in such bound states \cite{binding}.  
The trapping efficiency is very high. 
The theoretical modeling that has been done assumes ``rigid'' extended nuclei
with or without repulsive barriers, and some relativistic calculations have
also been carried out.  Electrons can bind to monopoles
 in a total energy zero state;  this probably
produces  a small mobile system which will transfer to a nuclear 
magnetic moment leaving it bound to a fixed nuclear site and   
permanently trapped.  Thus we assume all monopoles bind to appropriate
nuclei, i.e., those whose nuclear gyromagnetic ratio
is sufficiently large (anomalous).  These models predict, as summarized
in Ref.~\cite{binding}, that binding should
occur for $^{27}$Al (100\% natural abundance) and $^{207}$Pb (22\% natural) 
but not for $^9$Be (100\% natural).  However, the estimated 
binding energies, e.g., 0.5--2.5 MeV for aluminum, 
are large and comparable to shell model splittings, so we believe that
in the presence of the monopole the nucleus will undergo nuclear rearrangement
and binding should in general result, even for $^9$Be. Even an
unreasonably small estimate for the binding energy of 1 eV would give a lifetime
of 10 yr \cite{binding}. We have, therefore,
good reason to believe that stopped
monopoles will be trapped by the magnetic moments of nuclei.

\section{Analysis of the Data}
\label{secIV}

\begin{figure}
\centering
\includegraphics[height=12cm]{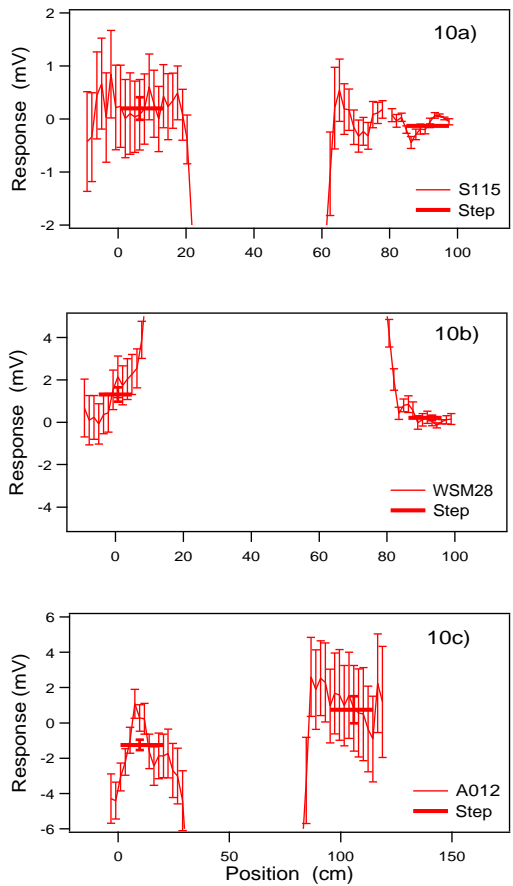}
\caption{Typical step plots: a) set 1, D0 aluminum, sample ``S115,''
b) set 2, CDF lead, sample ``WSM28,''
 and c) set 3, CDF aluminum, sample ``A012.''
The steps are $-0.4$, $-1.1$, and 2.0 mV, respectively.  
}
\label{fig10}
\end{figure}

The data analysis proceeded generally as follows, with modifications as
required for the different sets of samples.  The time sequences of the SQUID's
outputs were examined interactively, and bad sections,
primarily sections that contained a SQUID reset (see Sec.~\ref{secI}),
 deleted pairwise (to
keep induced signals canceling out pairwise) and the corresponding traces on 
the other SQUID also deleted (to keep them temporally consistent); 
typically 80--90\% of the traversals
remained.  A pedestal value, the SQUID output near the top end 
of each traversal (bottom for sample set 3),
 was subtracted from every voltage value
along that up (or down) traversal.  The values for each of some 90 (75 for
sample set 3) small ranges of vertical positions were averaged, removing
most of the random drift of the SQUIDs.  The two SQUIDs' data were averaged,
shifting one relative to the other by 10.1 cm in position in order to 
superimpose their dipole responses.  The background samples were analyzed
similarly and local groups of background runs were averaged.  These background
runs were subtracted from the samples' spectra.  A pair of horizontal lines 
was fit to two regions, one at the lower position and one at the upper.  The
difference in values of these two flat fits gives the step 
for that sample.  
Examples of processed data are shown, one for each sample set, in 
Fig.~\ref{fig10}.
The steps for each sample set are histogramed and analyzed for consistency
with the null hypothesis, absence of monopoles.  The background subtraction 
ensures that the distribution of steps centers on zero, since various 
background effects, such as the small effect of the magnetized string holding
the sample in set 1, has been removed. These analyses were performed with a 
large set of macro procedures written for the program WaveMetrics'
IGOR-Pro \cite{igor} running on Apple Macintosh computers.

For sample set 1 (D0 aluminum and beryllium) the background sequence was
one background sample (always the same sample) between every two samples
measured.  For sample set 2 (CDF lead) it was one background between every
three samples.  And for sample set 3, the other samples were used as 
backgrounds for any given sample.  The background averaging was done over
time blocks of data.  Each time the measurements were interrupted for
unmanned time, a liquid helium fill, a thunderstorm, etc., some possible 
change in the environmental background was likely, so these were considered 
time blocks that required an independent background subtraction.

Sample set 2's sample holder was a threaded metal rod with hexagonal nuts
clamping the rolled lead sample cylinder.  The overall magnetic dipole
background was generally dominated by the sample holder, making the analysis
of the CDF lead samples a bit harder than those of sample set 1.

Sample set 3 differed most in its procedures.  A
greater signal sensitivity should give a better overall signal-to-noise ratio
for each SQUID's data. As a major part of the noise was external to the SQUID
electronics, the overall gain of the controllers was increased 
(see Section \ref{secII} above)
by  doubling feedback resistor values and halving capacitor 
values of the final operational amplifiers.  
This sensitivity gain apparently would have occurred.  
However, as the experiment was winding down, with time, funding and manpower
coming to an end, the failure of one of the two SQUIDs for set 3 was not
repaired (in consideration of the time that would have been lost to warm-up, 
repair, cool-down, etc.).
The remaining SQUID had a small 0.4 Hz oscillation which we then dealt with
in the analysis software, by smoothing the spectrum over a number of the 10 Hz
data input values before editing the traversals. In addition, we extended
the range of the traversals to 1.2 meter, at a larger stepper motor interval, 
leading to fewer (the 75 bins mentioned above) 
vertical position bins.  The IGOR macros
were modified to allow the computer to delete pairs of traces based on a
chi-squared test of the consistency of differing traces in a sample's 
traversals from the average of those minimally edited (for bad traces, resets,
etc.).  Chi-squareds per degree of freedom greater 
than two were rejected, corresponding to a ten percent loss of traces on 
average.  

\section{Monopole Cross-Sections and Mass Limits}
\label{secV}
\begin{figure}
\centering
\includegraphics[height=4cm]{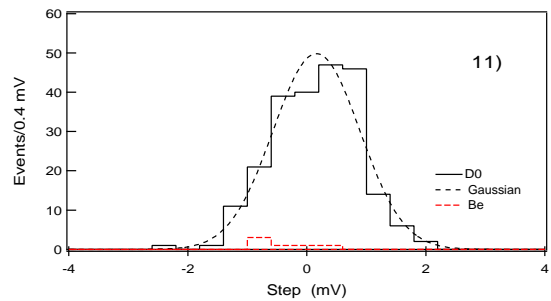}
\caption{Histogram of the steps from 222 aluminum and 6 beryllium samples
from the D0 experiment.  The total of 228 samples, with an rms spread
of 0.73 mV, is compared to a Gaussian with that same standard deviation.
}
\label{fig11}
\end{figure}

\begin{figure}
\centering
\includegraphics[height=12cm]{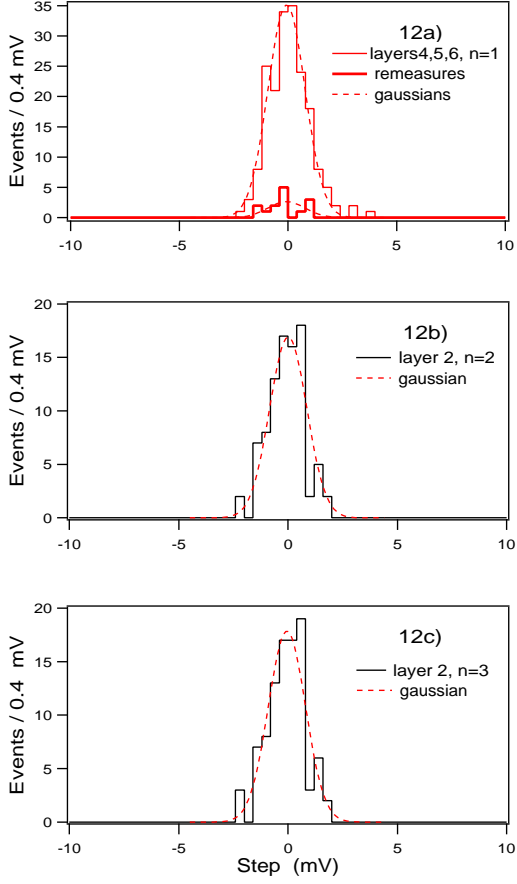}
\caption{Histograms of the steps for the CDF lead samples, with an rms 
dispersion
of 0.85 mV. a) 187 layer 4, 5, 6 samples along with the 14 remeasurements,
having $sS < 0.3$ mV and $dS < 0.45$ mV,
b) 90 layer 2 samples having $sS < 0.4$ mV and $dS < 0.6$ mV, and 
c) 95 layer 2
samples having $sS < 0.6$ mV and $dS < 0.8$ mV.  Also shown are the four
corresponding Gaussians for comparison.  Note that the distributions in a), b),
c) are used for limits on magnetic charges of $|n|=1,2,3$,  respectively.
}
\label{fig12}
\end{figure}

\begin{figure}
\centering
\includegraphics[height=4cm]{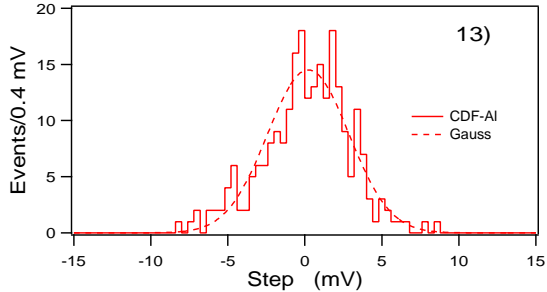}
\caption{Histogram of the steps from 240 aluminum samples ($sS<1.0$ mV)
from the CDF experiment, with an rms dispersion
 of 2.7 mV (reflecting a higher gain setting of SQUID DC1R, see
text) compared to the appropriate Gaussian.  These data are used for limits for
the three magnetic charge values $|n|=2,3$,  and 6. 
}
\label{fig13}
\end{figure}

The analyses of Sec.~\ref{secIV} above yield a number of ``step'' 
histograms.  Those
for set 1, D0 aluminum and beryllium, are shown in Fig.~\ref{fig11},
for set 2,
CDF lead, in Fig.~\ref{fig12},
 and for set 3, the CDF aluminum samples, 
in Fig.~\ref{fig13}
after taking account of the cuts on the errors discussed below.

\begin{figure}
\centering
\includegraphics[height=12cm]{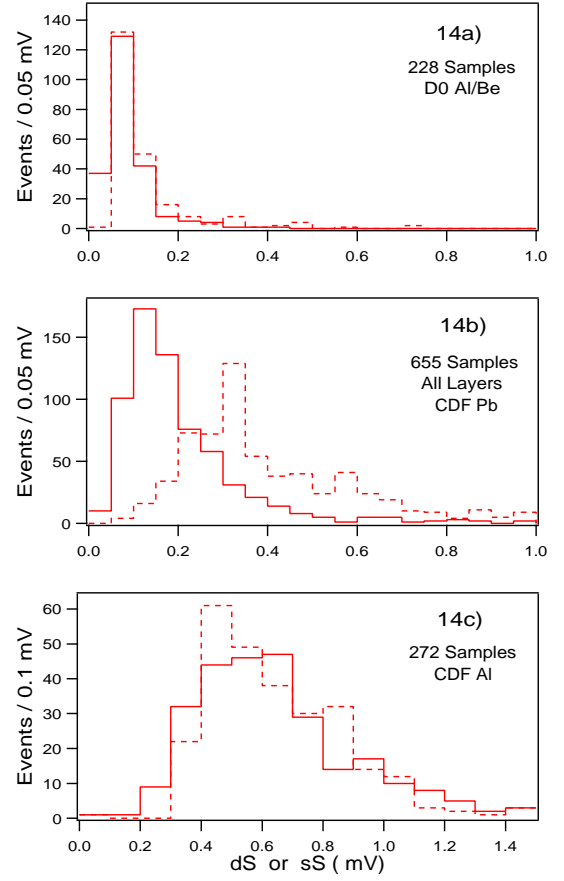}
\caption{The error histograms for $dS$ (dashed) and $sS$ (solid) for sample 
sets 1, 2 and 3 (remember the higher gain setting for set 3).  Note that the 
the individual $(dS, sS)$ error values for the step plots a, b, c, shown 
in Figure \ref{fig10}, are (0.22, 0.08), (0.34, 0.32) and (0.82, 0.48) mV, respectively.
}
\label{fig14}
\end{figure}

In these plots
the variable $dS$ is the standard statistical error on the step $S$, 
whereas $sS$ is
a systematic error which is the rms of the deviations of the actual step
from the flat fitted steps shown, e.g., in Fig.~\ref{fig10}.  
If the distribution
is flat (with little slope) $sS$ is small, whereas if it is badly sloped it is
large.  Bad measurements, e.g., very large dipole tails, etc., cannot give a
true measure of the step $S$, and so are removed by this cut.  Note that 
Fig.~\ref{fig10}b has a wide dipole response, 
but still an acceptable $sS$; its wide 
dipole response comes about because it is the superposition of two dipoles
separated in position.  
Some cases cut by a large $sS$ are due to larger or more
complicated superpositions of dipole responses.  
The histograms of $dS$ and $sS$ are shown in Fig.~\ref{fig14}: cuts on $dS$
and $sS$ are made in order to control the dispersion of the step histograms,
and losses of events are taken into account in the efficiency $\epsilon$
defined below.

A number of parameters need to be taken from these plots and interpreted to
yield the cross section and mass limits desired. 
Each plot determines an upper limit on the number ($N_{\rm ul}$) 
of monopoles of a given
magnetic charge ($n= 1,2,3,6$) for each sample set which in turn determines
an upper limit to the corresponding cross-section
\begin{equation}
\sigma <  \frac{N_{\rm ul}}{ \varepsilon \epsilon A \mathcal{L}},
\label{e14}
\end{equation}
where	$\varepsilon$ is the efficiency for the chosen signal 
(a monopole with charge $n$) to lie outside a cut excluding smaller
values of $|n|$,
$\epsilon$ is the efficiency of the sample set to cover the solid-angle
region chosen  and to correct for $dS$ and $sS$ cut losses,
and to account for losses of sample material.
$\mathcal{L}$ is the total luminosity for the $\overline p$-$p$ 
exposure delivered  ($172 \pm 8 \mbox{pb}^{-1}$ for D0 \cite{dolum} and 
$180 \mbox{pb}^{-1}$ $\pm 5$\% for CDF \cite{luminosity}).

The Drell-Yan modelling is that \cite{Kalbfleisch:2000iz} 
$d\sigma/dM_{\rm pair}\sim(\beta/M_{\rm pair})^3$, 
where $\beta$ is the velocity of the monopole in the rest
frame of the monopole-antimonopole pair.  This basic cross section,
involving the interaction suppression factor of $\beta^2$, is
multiplied by the phase space $\beta$
and convolved with the CETQ5 \cite{Lai:1999wy} quark-parton distributions
(PDFs). In Ref.~\cite{Kalbfleisch:2000iz} we normalized the integral of
this cross section to $(g/e)^2\sim4700n^2$ times the experimental CDF and D0
$p\bar p\to\mu\bar\mu$ cross sections 
\cite{Abe:1994kq,Abbott:1998rr,Kalbfleisch:2000iz} and obtained the $\sigma(M
\bar M)$ Drell-Yan cross section.  For Luo's thesis \cite{Luo:2002tm}
an a priori theoretical normalization was made, by inserting the quark-antiquark
to mu pairs cross section,
\begin{equation}
\sigma(q\bar q)=\frac{4\pi\alpha^2e_q^2}{3Q^2},\label{xx}
\end{equation}
where $e_q$ is the quark charge and $Q$ is the mass of the quark pair,
 in place of $d\sigma/dM$ above.  The summing over quark charges and convolving
 with PDFs, etc., was also done.  This procedure agrees with the first one
 within a factor of two, and is stable with respect to choice of PDF
 (e.g., MRST \cite{Martin:1998sq}).  Equation (\ref{xx}) has integrated out
 a $1+\cos^2\theta$ center-of-mass angular distribution appropriate to a $\mu
 \bar \mu$ virtual photon vertex.  
 The Drell-Yan Monte Carlo puts this factor back into 
 the equation and throws weighted events according to the mass and angular
 distributions of the model.  The thrown monopole events are Lorentz transformed
 to the laboratory frame and tracked through the various structural materials
 into the appropriate sample(s).  We also take into account the acceleration
 or deceleration due to any external magnetic fields present
 ($\sim30n$ GeV/m for CDF).  Those events
 entering and stopping in the sample layer(s) are summed into the ``accepted''
 cross section $\sigma_A$.  The acceptance is then $A=\sigma_A/\sigma$, where
 $\sigma$ is the total cross section without any path length or angle cuts.

\begin{figure}
\centering
\includegraphics[height=4cm]{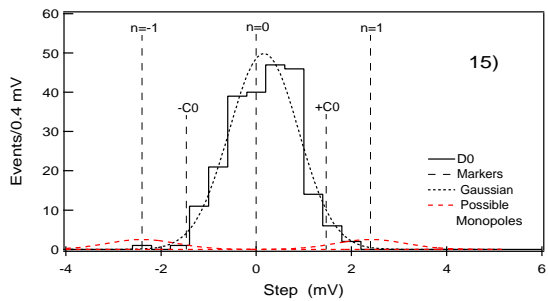}
\caption{A schematic of the upper limit number analysis parameters needed
to obtain the cross-section limits via Tables \ref{tabV} and \ref{tabVI}.
}
\label{fig15}
\end{figure}

In order to extract $N_{\rm ul}$, we must make cuts on the step histograms,
as shown in
Fig.~\ref{fig15}.
  We define a region about zero using $C_0$ (i.e., $0\pm C_0$)
 such that 
most (zero) steps lie there, leaving most of the potential 
$n \ne0$ charges lying outside
of this region.  The $C_0$ to $\infty$ and $-\infty$ to $-C_0$ 
region defines truncated Gaussians
for magnetic charge $\pm n$, whose standard deviation is
assumed (in the absence of further information) to be
the same as that of the central (zero) peak, and shown as the small
offset Gaussians in the figure.
These are centered upon the expected positions of the
charge $n$ monopoles, which is $n$ times the pseudopole 
calibration value from
Table \ref{tabII}, Sec.~\ref{secII}.  Then $\varepsilon$ 
is the fractional area 
of these truncated Gaussians between $C_0$ and infinity.  Also, the expected
number ($N_{\rm exp}$) of null events
outside $\pm C_0$ is the fraction of the central Gaussian lying outside times 
the total number of events in the histogram.\footnote{The value of $C_0$ is
chosen to minimize $N_{\rm exp}$ while retaining good efficiency for finding
$n\ne0$.}
 This number is to be
 compared to the observed 
number ($N_{\rm obs}$) lying outside.  We use the Feldman-Cousins tables
of $N_{\rm exp}$ vs $N_{\rm obs}$ to find $N_{\rm ul}$ \cite{cousins}
at 90\% confidence level.
These parameters are compiled and 
displayed in Tables \ref{tabV} and \ref{tabVI}, 
from which $\sigma$ is calculated and tabulated
(the lower limit on the monopole mass, $m_{\rm LL}$, is discussed below).

\begin{table*}
\caption{\label{tabV}Summary of parameters for the various sample 
sets and the various
magnetic charge values ($n$) of the step histograms needed via 
Fig.~\protect\ref{fig15}. Here ``response'' refers to the mean value
expected for a monopole of strength $n$.  The table lists only those
samples that have sufficient sensitivity for that value of $n$.}
\begin{ruledtabular}
\begin{tabular}{crcccccc}
Set&$n$&Dispersion&$C_0$&Response&
$N$&$N_{\rm exp}$&N$_{\rm obs}$\\
&&(mV)&(mV)&(mV)&&&\\
1 Al&1&0.73&1.47&2.46&222&9.8&8\\
1 Al RM&1&0.73&1.47&2.46&8&0.4&0\\
2 Pb&1&0.85&1.6&2.46&187&11.2&14\\
2 Pb RM&1&0.85&1.6&2.36&14&0.8&0\\
1 Al&2&0.73&3.0&4.92&222&0&0\\
2 Pb&2&0.85&3.5&4.92&90&0&0\\
3 Al&2&2.7&8.8&10.64&240&0.3&0\\
1 Be&3&0.73&3.0&7.38&6&0&0\\
2 Pb&3&0.85&3.5&7.38&95&0&0\\
3 Al&3&2.7&8.8&16.0&240&0.3&0\\
1 Be&6&0.73&3.0&14.8&6&0&0\\
3 Al&6&2.7&8.8&31.9&240&0.3&0\\
\end{tabular}
\end{ruledtabular}
\end{table*}

\begin{table*}
\caption{\label{tabVI}The remaining parameters and final limits 
(90\% confidence level) from 
Drell-Yan modeling for cross-sections and monopole masses.  The upper
limit $N_{\rm ul}$ comes from a Feldman-Cousins analysis \cite{cousins}.
Note: Set I Al RM and 2 Pb RM (lines 2 and 4) are corrected for losses of samples
not remeasured by dividing by $\varepsilon$ again.}
\begin{ruledtabular}
\begin{tabular}{ccccccccc}
Set&$n$&$N_{\rm ul}$&$\varepsilon$&$\epsilon$&$A_{+1}$&$\mathcal{L}$
&$\sigma^{\rm ul}_{+1}$&$m^{\rm LL}_{+1}$\\
&&(90\% CL)&&&&(pb$^{-1}$)&(pb)&(GeV/$c^2$)\\
1 Al&1&4.4&0.91&0.94&0.026&172&1.2&250\\
1 Al RM&1&2.0&0.91&0.94&0.026&172&0.6&275\\
2 Pb&1&10.3&0.84&0.53&0.013&180&9.9&180\\
2 Pb RM&1&1.7&0.82&0.53&0.011&180&2.4&225\\
1 Al&2&2.4&1.00&0.94&0.007&172&2.1&280\\
2 Pb&2&2.4&0.95&0.83&0.017&180&1.0&305\\
3 Al &2&2.1&0.75&0.88&0.10&180&0.2&365\\
1 Be&3&2.4&1.00&1.0&0.0036&172&3.9&285\\
2 Pb&3&2.4&1.00&0.87&0.028&180&0.5&350\\
3 Al&3&2.1&0.997&0.88&0.19&180&0.07&420\\
1 Be&6&2.4&1.00&1.0&0.013&172&1.1&330\\
3 Al&6&2.1&1.00&0.88&0.057&180&0.2&380\\
\end{tabular}
\end{ruledtabular}
\end{table*}

 The only non-standard 
cases are the remeasurements of sets 1 and 2, in which the loss of possible 
monopole events not remeasured 
has to be taken into account, by another division by $\varepsilon$, 
as remarked in 
Table \ref{tabVI}.  We note that the cross-section limits range from 0.07 to 
9.9 pb,
a considerable improvement over earlier results from other experiments.  

The above results are based on an model using a $1 + \cos^2\theta$
angular distribution, appropriate for the coupling of spin-1/2 leptons to
photons.  As the angular distribution of magnetic monopoles is unknown,
we try also two other distributions to give a flavor of the changes that
would occur:  isotropic (i.e.,  1) and $\sin^2\theta=1 - \cos^2\theta$.  These 
alternative possibilities are given in Table \ref{tabVII}.  The total Drell-Yan
cross-sections for these three angular distributions are in the ratio
of 1 : 3/4 : 1/2 respectively when interpreting as mass limits.
\begin{table*}
\caption{\label{tabVII} Alternative interpretations for different production 
angular
distributions of the monopoles, comparing $1$ and $1-\cos^2\theta$ to the
(repeated here) $1+\cos^2\theta$ limits.  Here the acceptance $A_a$
corresponds to the distribution $1+a\cos^2\theta$, and similarly for the
cross section and mass limits (all at 90\% confidence level).}
\begin{ruledtabular}
\begin{tabular}{cccccccccc}
Set&$n$&$\sigma^{\rm ul}_{+1}$&$m^{\rm LL}_{+1}$&$A_0$&
$\sigma^{\rm ul}_{0}$&$m^{\rm LL}_{0}$&$A_{-1}$&
$\sigma^{\rm ul}_{-1}$&$m^{\rm LL}_{-1}$\\
&&(pb)&(GeV/$c^2$)&&(pb)&(GeV/$c^2$)&&(pb)&(GeV/$c^2$)\\
1 Al&1&1.2&250&0.024&1.2&240&0.021&1.4&220\\
1 Al RM&1&0.6&275&0.024&0.6&265&0.021&0.7&245\\
2 Pb&1&9.9&180&0.011&12&165&0.0055&23&135\\
2 Pb RM&1&2.4&225&0.009&2.9&210&0.0045&5.9&175\\
1 Al&2&2.1&280&0.0068&2.2&270&0.0060&2.5&250\\
2 Pb&2&1.0&305&0.018&0.9&295&0.016&1.1&280\\
3 Al &2&0.2&365&0.10&0.2&355&0.096&0.2&340\\
1 Be&3&3.9&285&0.0025&5.6&265&0.0003&47&180\\
2 Pb&3&0.5&350&0.029&0.5&345&0.031&0.5&330\\
3 Al&3&0.07&420&0.20&0.07&410&0.24&0.06&405\\
1 Be&6&1.1&330&0.008&1.7&305&0.0008&18&210\\
3 Al&6&0.2&380&0.066&0.2&375&0.082&0.2&370\\
\end{tabular}
\end{ruledtabular}
\end{table*}

\begin{figure}
\centering
\includegraphics[height=13cm]{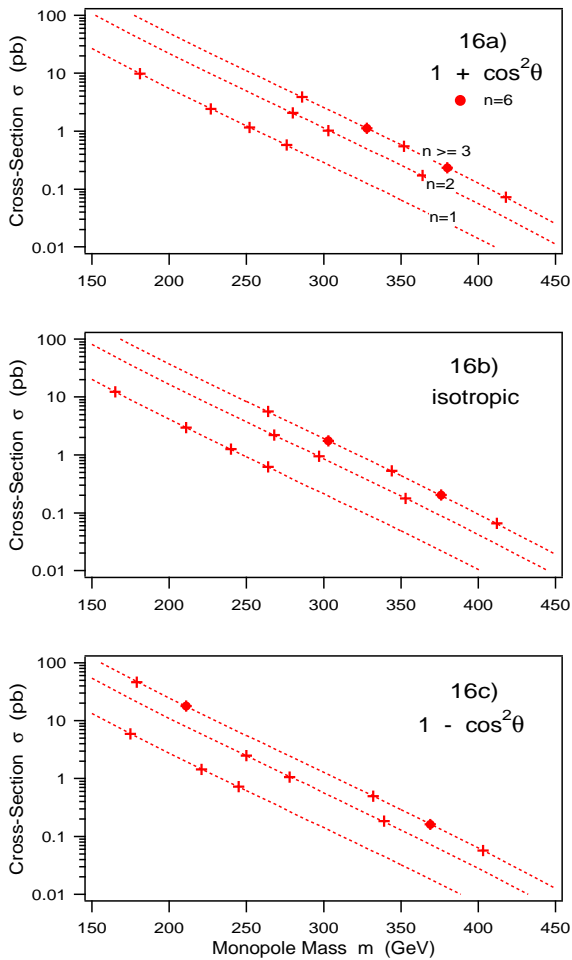}
\caption{
 The curves are Drell-Yan cross sections versus monopole mass with
cross-section upper limits (90\% CL)
interpreted as mass limits (cross section upper
limit intersects Drell-Yan curve at estimated lower limit of mass as shown
by $+$ and $\bullet$ markers).  Three possible center-of-mass angular distributions are
considered, of form  $1 + a\cos^2\theta$, with $a= 1$, 0 and $-1$, 
 respectively.}
\label{fig16}
\end{figure}

The cross-section limits above are model dependent, but only moderately so,
since only the shape of the $d\sigma/dM_{\rm pair}$ 
distributions is relevant to the
acceptance $A$, apart from the straightforward angular distribution,
$dE/dx$, and magnetic field tracking 
considerations.  The interpretation of these limits as monopole mass limits
is directly model dependent. We have used the Drell-Yan model for lack of any
field theoretic results. (For the status of the theory, see 
Ref.~\cite{Gamberg:1999hq}.)
 We modify the couplings from $e^2/\hbar c$ to 
$g^2/\hbar c$ (as appropriate in $dE/dx$ calculations)
rather naively, but we have also included $\beta^3$ velocity suppression
factors, a conservative choice.  Moreover,
we note that there is a unitarity limit that comes in at $n\approx 3$, and we
use $n=3$ cross-sections as the unitarity limit for all $n\ge3$  (i.e., 
we use three
cross-sections, those for $n=1$ and $n=2$, and that for $n=3$ for higher $n$).
  Then,
taking the intersection of the $\sigma$ limits with the Drell-Yan curves we
obtain the lower limit for the mass
(to the nearest 5 GeV) of each given monopole case, as entered 
in Tables \ref{tabVI} and \ref{tabVII}, and shown in Fig.~\ref{fig16}.  
These mass limits are an improvement of a factor of
about two or more over previous direct experimental results.

\section{Summary}
\label{secVI}

We have improved the cross-section and  mass limits on directly produced
magnetic monopoles over the best previous other limits of Bertani et 
al.~\cite{Bertani:1990tq} and the lunar rock limits of Ross et 
al.~\cite{Ross:1973it}.  We have
improved and extended our own prior limits from D0 samples 
\cite{Kalbfleisch:2000iz} to those from CDF, with a consistent overall 
analysis, presented in this paper.

Since limits are given, it is hard to assign errors.  However, we can give a 
range or ``spread'' of the values from the quoted means to give a flavor of the
variations to be expected.  These spreads are in general a combination of
statistical and systematic errors.  Since we are modelling according to the
Drell-Yan process, the systematic errors on the acceptances are unknown; it
would take the evaluation of another reasonable proposed model, or a true
quantum field theoretic prediction, to assess that.
For example, a factor of two change in the cross section
would lead to a change in the mass limits of about 20 GeV.  The acceptances are 
estimates in any case, since the modeling of the detector elements, the 
$dE/dx$ losses, and the tracking of the particles through the various elements
are not exact.  The statistical error on the acceptances are 3\% individually
and 1\% from smoothing over a mass interval (with a quadratic functional fit)
of some 9--11 calculations spanning a 200--300 GeV mass region.  The 
efficiencies ($\varepsilon$ and $\epsilon$) are good to about 2\% and the
luminosities to some 5\%.  Combining 3\%, 2\%, 2\%, and 5\% quadratically gives
a spread of $\sim\pm7$\% on $\sigma_{\rm ul}$ in addition to the variations
due to differing assumed angular distributions.  Since the Dress-Yan cross
sections are closely linear on a semilog plot against mass, this converts to a
$\sim\pm1$ GeV spread in mass.  The unknown angular distributions give a 
greater uncertainty.

Since the true angular distributions are unknown, we take as a ``base'' an
isotropic one (at 90\% CL).  
In addition, we choose the largest mass limit for each 
value of the magnetic charge $n$.  These come from D0 Al set 1 for $n=1$, 
and CDF Al set 3 for $n=2, 3$ and 6. From Fig.~\ref{fig16} and Table 
\ref{tabVII}, we see that these have cross section limits of 0.6, 0.2,
0.07, and 0.2 pb, respectively.  The corresponding mass limits are
265, 355, 410, and 375 GeV, with spreads of ($+10,-20$), ($+10,-15$), 
($+10,-5$),
and $\pm5$, respectively, due to replacing isotropic distributions by
ones of the form $1\pm\cos^2\theta$.

These cross section limits are some 250 to 2500 times smaller and the mass
limits are about 3 times larger than those of Bertani et 
al.~\cite{Bertani:1990tq}, and a
significant improvement over those of Ross et al.~\cite{Ross:1973it}, as well.

\begin{acknowledgments}
We acknowledge the support of the U.S. Department of Energy under Grants
DE-FG02-95ER40923 and DE-FG03-98ER41066.  We thank the University of 
Oklahoma's (OU's) Departments of Physics and Astronomy and of Aerospace
and Mechanical Engineering for support and laboratory space. We thank J.
Young, B. Bergeron and R. Littel of OU's Physics Instrument Shop and
A. Wade of OU's Electronics Shop for all their efforts.  Questions
regarding the SQUIDs were answered by D. Polancic of
Quantum Design.  Also thanks to our
colleagues S. Murphy and L. Gamberg of OU, M. Longo of the University of 
Michigan, T. Nicols, M. Kuchnir, and H. Haggerty of Fermilab for their various
contributions.  We thank Fermilab and the D0 and CDF Collaborations for the
samples.  In addition, we acknowledge the operational and analysis work at
various times
of a number of graduate and undergraduate students that allowed us to carry
out the measurements reported in this paper: I. Hall, C. Hladik, T. Zheng,
D. Abraham, R. Abraham, W. Bullington, Y. Milton, S. Miyashita, M. Nguyen,
B. Schlecht and D. Stewart.
\end{acknowledgments}

\appendix*
\section{Simplified Theory of Monopole Detector}
\begin{figure}
\centering
\begin{picture}(200,200)
\put(50,0){\line(0,1){200}}
\put(150,0){\line(0,1){200}}
\qbezier[80](100,0)(100,100)(100,200)
\qbezier(50,0)(100,10)(150,0)
\qbezier(50,0)(100,-10)(150,0)
\put(100,20){\makebox(0,0){$\bullet$}}
\put(110,20){\makebox(0,0){$g$}}
\put(25,20){\makebox(0,0){$z'$}}
\qbezier(75,40)(100,50)(125,40)
\qbezier(75,40)(100,30)(125,40)
\put(25,40){\makebox(0,0){$Z$}}
\put(100,40){\vector(1,0){25}}
\put(130,40){\makebox(0,0){$r$}}
\put(100,100){\vector(1,0){50}}
\put(125,90){\makebox(0,0){$a$}}
\put(25,100){\makebox(0,0){$z$}}
\put(25,120){\makebox(0,0){$\uparrow$}}
\put(25,0){\makebox(0,0){$z=0$}}
\qbezier[50](50,200)(100,210)(150,200)
\qbezier[50](50,200)(100,190)(150,200)
\end{picture}
\caption{Diagram of monopole detector}
\label{figapp1}
\end{figure}
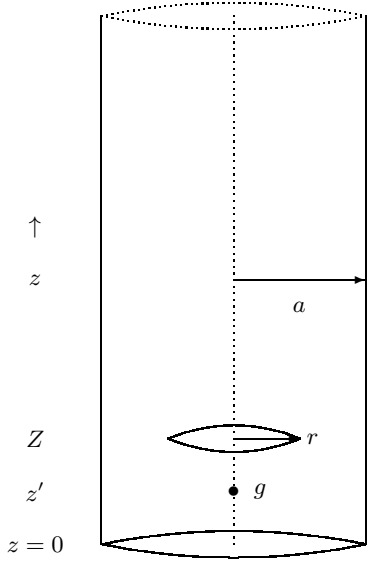

This appendix describes the basis of the functioning of our magnetic monopole
dectector.  It works by detecting the magnetic flux intercepted by
a superconducting loop contained within a superconducting cylinder.
The detector is sketched in Fig.~\ref{figapp1}.

In order to incorporate finite-size effects,
we consider first a perfectly conducting right circular cylinder of radius 
$a$ of semi-infinite length, with axis along the $z$-axis, and with a
perfectly conducting circular bottom cap at $z=0$.  We use cylindrical
coordinates $\rho$, $\theta$, and $z$.

Because the boundaries are superconductors, the normal component of
$\bf B$ must vanish on the surfaces, that is,
\begin{equation}
B_\rho\bigg|_{{\rho=a}\atop{z>0}}=0,\quad B_z\bigg|_{z=0}=0.
\label{a0}
\end{equation}

Now suppose a magnetic pole of strength $g$ is placed on the
$z$ axis at $z=z'>0$.  This could either be a magnetic monopole (magnetic
charge) or one pole of a very long electromagnet (``pseudopole'').
Imagine a circular conducting loop of radius $r<a$ centered on the axis
of the cylinder and perpendicular to that axis, with center at $z=Z$.
Inside the cylinder and outside of the loop  $\bf B$ is
derivable from a magnetic scalar potential,
\begin{equation}
{\bf B}=-\mbox{\boldmath{$\nabla$}}\phi_M,
\label{a1}
\end{equation}
since we may ignore the displacement current, because the
time variation is negligible.   $\phi_M$ satisfies
Poisson's equation, in cylindrical coordinates:
\begin{eqnarray}
\mbox{\boldmath{$\nabla$}}\cdot\mathbf{B}&=&
-\left(\frac1\rho\frac\partial{\partial \rho}\rho\frac\partial{\partial \rho}
+\frac1{\rho^2}\frac{\partial^2}{\partial\theta^2}+\frac{\partial^2}
{\partial z^2}\right)\phi_M\nonumber\\
&=&4\pi g\delta({\bf r-r'}),
\label{a2}
\end{eqnarray}
where $\mathbf{r}'$ is the position of the monopole, $\mathbf{r'}=(\rho',
\theta',z')$.
This is the equation for a Green's function, which we can express
in separated variables form.  That is, we write
\begin{eqnarray}
\phi_M&=&\frac2\pi\int_0^\infty dk\cos kz\cos kz'\nonumber\\
&&\quad\times\sum_{m=-\infty}^\infty
\frac{1}{2\pi}e^{im(\theta-\theta')}g_m(\rho,\rho';k),
\label{a4}
\end{eqnarray}
where, in view of the first boundary condition in Eq.~(\ref{a0}),
we may express the reduced Green's function in terms of modified Bessel 
functions:
\begin{eqnarray}
g_m(\rho,\rho';k)&=&-4\pi gI_m(k\rho_<)\nonumber\\
&&\quad\times\left[K_m(k\rho_>)-I_m(k\rho_>)
\frac{K'_m(ka)}{I'_m(ka)}\right],\nonumber\\
\label{a5}
\end{eqnarray}
where $\rho_<$ ($\rho_>$) is the lesser (greater) of $\rho$, $\rho'$.
If the monopole is confined to the $z$ axis, only the $m=0$ term survives:
\begin{eqnarray}
\phi_M&=&-\frac{4g}{\pi}\int_0^\infty dk\cos kz\cos kz'\nonumber\\
&&\quad\times\left[K_0(k\rho)
+I_0(k\rho)\frac{K_1(ka)}{I_1(ka)}\right],
\label{a6}
\end{eqnarray}
which uses
\begin{equation}
I_0'(x)=I_1(x),\quad K_0'(x)=-K_1(x).
\label{a7}
\end{equation}

By integrating over the cross section of the loop using
\begin{subequations}
\begin{eqnarray}
\int_0^x dt\,t\,K_0(t)&=&-x\,K_1(x)+1,\\
\int_0^x dt\,t \,I_0(t)&=&x\,I_1(x),
\label{a8}
\end{eqnarray}
\end{subequations}
we obtain the following formula for the magnetic flux subtended
by the loop,
\begin{equation}
\Phi=\int d{\bf S\cdot B}=4\pi g\left[\eta(Z-z')-F(Z,z')
\right],\label{a9}
\end{equation}
where the step function is
\begin{equation}
\eta(x)=\left\{\begin{array}{cc}
1,&x>0,\\
0,&x<0,\label{a10}
\end{array}\right.
\end{equation}
and the response function is
\begin{eqnarray}
F(z,z')&=&\frac2\pi\frac{r}{a}\int_0^\infty dx\,\sin x\frac{z}{a}
\cos x\frac{z'}{a}\nonumber\\
&&\times\left\{K_1(xr/a)-I_1(xr/a)\frac{K_1(x)}{I_1(x)}\right\},\nonumber\\
\label{a11a}
\end{eqnarray}

Now suppose that the pole is {\em slowly\/} moved from 
a point far above the loop, $z'=+\infty$, to a point below the loop,
$z'=z_0$, $Z>z_0$.  Then from Maxwell's equation
\begin{equation}
\mbox{\boldmath{$\nabla$}}\times{\bf E}=-\frac1c\frac{\partial}{\partial t}
{\bf B}-\frac{4\pi}c{\bf J}_m,
\label{a11}
\end{equation}
where $\mathbf{J}_m$ is the magnetic current density,
the emf induced in the loop is
\begin{equation}
{\cal E}=\oint{\bf E}\cdot d{\bf l},
=-\frac{d\Phi}{cdt}+\frac{4\pi}{c}g\delta(t),
\label{a13}
\end{equation}
if $t=0$ is the time at which the pole passes through the plane of the
loop.  The net change in emf gives rise to a persistent current $I$ in the
{\em super}conducting loop,
\begin{eqnarray}
LI=\int_{-\infty}^\infty {\cal E} dt=-\frac1c\Delta\Phi+\frac{4\pi}c g
=\frac{4\pi}c gF(Z,z_0),
\label{a14}
\end{eqnarray}
where $L$ is the inductance of the loop, and the response function
$F$ is given in Eq.~(\ref{a11a}).  This is just a statement of the Meissner
effect, that the flux change caused by the moving monopole is cancelled by
that due to the current set up in the loop.

Now suppose that the loop is very far from the bottom cap,
$Z\gg a$.  Then only small $x$ contributes to the integral in Eq.~(\ref{a11a}),
and  it is easy to see that
\begin{equation}
\int_{-\infty}^\infty {\cal E}dt= \frac{4\pi g}c\left(1-\frac{r^2}{a^2}\right),
\label{a15}
\end{equation}
so the signal is maximized by making the loop as small as possible, relative
to the radius of the cylinder.  We get the full flux of the monopole
only for a loop in empty space, $a/r\to\infty$.
This perhaps counterintuitive effect is
due to the fact that the superconducting walls confine 
the magnetic flux to the interior of the cylinder.
Thus for the superconducting can, the induced current in the detection loop
caused by the passage of a monopole from $z'=\infty$ to $z'=0$ is
\begin{equation}
LI=\frac{4\pi g}c-\frac{\Delta\Phi}c=\frac{4\pi g}c-\frac{\Phi(z'=0)}c,
\label{a16}
\end{equation}
which yields the result (\ref{a15}) 
if one assumes that the magnetic field is uniform across the can's cross
section at the position of the loop when the pole is at the bottom, because
all the flux must pass up through the can.  If we consider, instead, an
infinite, open-ended, superconducting cylinder, with the monopole passing
from $z=+\infty$ to $z=-\infty$, at either extreme half the flux must
cross the plane of the loop, so with the uniformity assumption we get the
same result:
\begin{equation}
LI=\frac{4\pi g}c-\frac{\Delta\Phi}c
=\frac{4\pi g}{c}\left(1-\frac{r^2}{a^2}\right).
\label{a17}
\end{equation}
The simple assumption of a uniform magnetic field is apparently justified
by the exact result (\ref{a15}).

We conclude this appendix by noting how the exact calculation is modified
for an infinite superconducting cylinder.  In the magnetic scalar
potential, the integral over $k$ mode functions in Eq.~(\ref{a4}) is replaced by
\begin{equation}
\int_{-\infty}^\infty\frac{dk}{2\pi}e^{ik(z-z')},
\label{a18}
\end{equation}
which has the effect of replacing the flux expression (\ref{a9})
by
\begin{equation}
\Phi=2\pi g[\epsilon(z-z')-F(Z-z',0)],
\label{a19}
\end{equation}
where
\begin{equation}
\epsilon(x)=\left\{\begin{array}{cc}
1,&x>0,\\
-1,&x<0.\end{array}\right.
\end{equation}
Then the induced current in the detection loop when the monopole passes
from a point above the loop  $z'=Z+\xi$ to a point, equidistant, below
the loop, $z'=Z-\xi$, is
\begin{equation}
LI=\frac{4\pi g}cF(\xi,0)\to \frac{4\pi g}c\left(1-\frac{r^2}{a^2}\right),
\label{a20}
\end{equation}
where the last limit applies if $\xi/a\gg1$.  This result coincides with
that in Eq.~(\ref{a15}).
The function $R(\xi)=\frac12F(\xi,0)/(1-r^2/a^2)+\frac12$,
corresponding to a monopole starting from a point $z_1$ far above the loop,
$z_1-Z\gg a$, and ending at a point $z_0=Z-\xi$,
 is plotted as a function of $\xi$ 
for our parameter values in Fig.~\ref{fig7}c, where it is shown to
agree well with experimental data.  This response
function  coincides
with the result obtained from Eq.~(\ref{a14}), 
because
\begin{eqnarray}
F(Z,Z-\xi)&=&\frac12 F(2Z-\xi,0)+\frac12 F(\xi,0)\nonumber\\
&\approx&\frac12\left(1-\frac{r^2}{a^2}\right)+\frac12F(\xi,0),
\end{eqnarray}
if $Z/a\gg1$. This shows that
the effect of the endcap (which of course is not present in actual detector)
is negligible,  demonstrating that the fact that the superconducting
shield is of finite length is of no significance.

\end{document}